\documentstyle[12pt,aaspp4,flushrt,eqsecnum]{article}
\newcommand{\ev}{\mbox{\rm \,eV}}

\newcommand{\kev}{\mbox{\rm \,keV}}

\newcommand{\thte}{\theta_{\!e}}
\newcommand{\thtr}{\theta_{\!r}}
\newcommand{\np}{\frac{\partial n}{\partial x}}
\newcommand{\npp}[1]{\frac{\partial^{#1}n}{\partial x^{#1}}}

\newcommand{\eqref}[1]{(\ref{#1})}
\newcommand{\et}[1]{e^{\mbox{\footnotesize $#1$}}}
\newcommand{\bbeta}{\mbox{\boldmath $\beta$}}
\newcommand{\bp}{\mbox{\boldmath $p$}}
\newcommand{\bk}{\mbox{\boldmath $k$}}
\newcommand{\bphat}{\hat{\mbox{\boldmath $p$}}}
\newcommand{\bkhat}{\hat{\mbox{\boldmath $k$}}}
\newcommand{\bgrad}{\mbox{\boldmath $\nabla$}}
\newcommand{\bx}{\mbox{\boldmath $x$}}
\newcommand{\dt}{\! \cdot \!}
\begin{document}

\title{Comptonization of an isotropic distribution in moving media:
higher-order effects}
 
\author{Anthony Challinor\footnote{E-mail: A.D.Challinor@mrao.cam.ac.uk}
\&\ Anthony Lasenby\footnote{E-mail: A.N.Lasenby@mrao.cam.ac.uk}}
\affil{Astrophysics Group, Cavendish Laboratory,
Madingley Road, Cambridge CB3 0HE, UK.}

\begin{abstract}

We consider the Comptonization of an isotropic radiation field by a thermal
distribution of electrons with non-vanishing bulk velocity. We include
all relativistic effects, including induced scattering and electron
recoil, in the derivation of a kinetic equation which is correct
to $O(\thte^{2}, \beta\thte^{2}, \beta^{2}\thte)$, where $\beta$ is the
bulk velocity (in units of $c$) and $\thte$ is the ratio of the electron
temperature to mass.
The result given here manifestly conserves photon number, and easily
yields the energy transfer rate between the radiation and electrons. We also
confirm recent calculations of the relativistic corrections to the
thermal and kinematic Sunyaev-Zel'dovich effect.
\end{abstract}


\section{Introduction}

Many astrophysical processes depend on the Compton scattering of
photons by a thermal distribution of electrons. Examples from
cosmology include the
evolution of anisotropies in the cosmic microwave background (CMB) radiation
in the pre-recombination era~\cite{peebles70}, and the Sunyaev-Zel'dovich
distortion~\cite{zel69} in the CMB spectrum due to passage through hot
clusters. The Comptonization is described by the Boltzmann equation, which
is an integro-differential equation for the evolution of the photon
distribution function. For a detailed discussion of the Compton scattering
kernel, see Kershaw, Prasad, \&\ Beason (1986).
In situations of astrophysical interest, it is often the case
that the energy transfer during scattering is sufficiently small to allow
one to make a Fokker-Planck expansion of the scattering kernel. In this case,
the Boltzmann equation may be replaced by a hierarchy of differential equations
for the moments of the distribution function.

The most complete analytic treatment to date of Comptonization in a moving
media was given by Psaltis \&\ Lamb (1997), who derived the first two
moments of the photon kinetic equation for arbitrary (anisotropic)
photon and electron distributions.
This derivation was correct to first-order in $\hbar \omega /m_{e}c^{2}$
and $\thte\equiv k_{B} T_{e} /m_{e} c^{2}$, where $\omega$ is the photon
frequency, $T_{e}$ is the electron temperature and $m_{e}$ the electron mass,
and to second-order in $\beta$, the bulk velocity of the electrons in units of
$c$. Psaltis \&\ Lamb (1997) showed the importance of using the full
relativistic cross section for Compton scattering in order to obtain the
kinetic equation to the required accuracy. Although the generality of the
results in Psaltis \&\ Lamb (1997) ensures that they are sufficient to
describe many astrophysical processes, there are situations of interest
where the inclusion of higher-order relativistic corrections is
necessary. One such example is the Sunyaev-Zel'dovich effect for hot clusters,
where $k_{B} T_{e}$ may be as high as $15\kev$. In this example, the radiation
is initially isotropic in the cosmic frame, and the optical depth to Compton
scattering through the cluster is sufficiently small that the effect on the
scattering of anisotropies induced by
any peculiar velocity of the cluster may be neglected (the probability
of a photon undergoing multiple scattering is very low).
The effect of relativistic
corrections on the thermal Sunyaev-Zel'dovich effect ($\beta=0$) has
been studied numerically (Rephaeli 1995; Rephaeli \&\ Yankovitch 1997)
and analytically (Stebbins 1998; Challinor \&\ Lasenby 1998; Itoh,
Kohyama, \&\ Nozawa 1998). Recently, these analyses have been extended
to include the effects of any peculiar velocity of the cluster
(Nozawa, Itoh, \&\ Kohyama 1998; Sazonov \&\ Sunyaev 1998). It was shown
that for typical values of the peculiar velocity $\beta \simeq 1/300$, the
relativistic corrections to the kinematic effect are $\simeq 8\%$,
and arise from a term of $O(\beta \thte)$ which is not included in the
analysis of Psaltis \&\ Lamb (1997).

In the course of their derivation of the corrections to the kinematic
Sunyaev-Zel'dovich effect, Nozawa et al.\ (1998) described a covariant
method for performing a Fokker-Planck expansion of the photon
kinetic equation for $\beta\neq 0$ by extending the method used
in Challinor \&\ Lasenby (1998), where the electron bulk velocity
was neglected ($\beta = 0$). This method includes all relativistic effects,
including induced scattering and electron recoil in a unified manner.
However, Nozawa et al.\ did not go on to evaluate the detailed form
of the kinetic equation including all these effects; instead they gave only
the resulting equation for a Planck distribution of isotropic radiation
at temperature $T_{r}$ in the limit that $T_{r} \ll T_{e}$. This equation
is sufficient to describe the dominant corrections to the
Sunyaev-Zel'dovich effect (which was the purpose of their paper), but since
the effects of recoil and induced scattering do not enter in this limit,
their results are not sufficiently general to describe other interesting
properties of the Comptonization process, such as the energy transfer rate
between the electrons and (hot) radiation. In fact, the dominant
corrections to the Sunyaev-Zel'dovich effect have been derived independently
by Sazonov \&\ Sunyaev (1998) with recoil and induced scattering
neglected from the start. Their calculation is much simpler than that
in Nozawa et al.\ (1998), since they need only use the Thomson cross
section. In the present paper, which is intended to complement the paper by
Nozawa et al.\ (1998), we derive the detailed form of the photon
kinetic equation describing Comptonization of an initially isotropic radiation
field in moving media, in the limit of low optical depth. In this limit,
the Compton scattering term in the Boltzmann equation may be evaluated for
an isotropic distribution, and, since multiple scattering is very improbable,
the effects of the velocity-induced anisotropies on subsequent scatterings
can be safely ignored. We feel that this result, which is omitted from the
Nozawa et al.\ (1998) analysis, could be valuable to the astrophysics community
at large since Comptonization is central to many problems.
Relativistic effects may be fully included by a
systematic expansion in the parameters $\thte$ and $\beta$. The new result
given here is written in a form that manifestly conserves the number of
photons, and allows a simple calculation of the energy transfer rate between
the electrons and the radiation. In the limit of $\beta=0$ we recover the
expression given in Challinor \&\ Lasenby (1998). For $\beta\neq 0$,
our result yields corrections at higher order than given elsewhere
the literature. We give the kinetic equation correct to
$O(\thte^{2},\beta \thte^{2},\beta^{2}\thte)$, before calculating the
rate of energy transfer for a Planck distribution correct to
$O(\thte^{2},\thtr^{2},\beta\thte,\beta\thtr)$,
where $\thtr \equiv k_{B} T_{r}/m_{e} c^{2}$. We end with a brief discussion
of the terms in the kinetic equation that are required to describe the
Sunyaev-Zel'dovich effect, obtaining results in full agreement with
those in Nozawa et al.\ (1998).
We use units with $\hbar
=c = k_{B} = 1$ in the following, unless stated otherwise.

\section{Extending the Kompaneets equation}

We consider the Comptonization of an unpolarised, initially isotropic
radiation field by a thermal distribution of electrons at temperature
$T_{e}$ which has bulk velocity $\bbeta$ relative to the radiation.
The optical depth is assumed to be sufficiently low that the radiation
may be treated as isotropic throughout the interaction with the medium.
We shall work exclusively in the frame in which the radiation is
isotropic, but it should be emphasised that we express our final results in
terms of the electron number density \emph{in the rest frame of the
scattering medium}, $N_{e}$.
The electron distribution function is denoted by $f(E, \bphat)$
and the photon distribution by $n(\omega, \bkhat)$. (Note that we use
distribution functions normalised to equal the mode occupation numbers.)
Here, the electron
energy is $E$, and the direction of propagation $\bphat$. For
the photons, $\omega$ denotes the energy and $\bkhat$ the direction
of propagation. Neglecting the effects of electron degeneracy,
the Boltzmann equation for the evolution of
$n(\omega, \bkhat)$ may be written as~\cite{buchler76}
\begin{eqnarray}
\lefteqn{%
\frac{Dn(\omega, \bkhat)}{D t} = -2 \int
\frac{d^{3} \bp}{(2\pi)^{3}} d^{3} \bp' d^{3} \bk'
W \Big\{ n(\omega ,\bkhat) [1+ n(\omega', \bkhat')]
f(E, \bphat)}\phantom{xxxxxxxxxxxxxxxxxxxxxxxxxx}
\nonumber \\
&& - n(\omega', \bkhat')[1+n(\omega, \bkhat)] f(E', \bphat')
\Bigl\},
\label{eq_bol}
\end{eqnarray}
where the operator $D/Dt$ denotes $\partial_{t} + \bkhat \dt \bgrad$.
The invariant transition amplitude $W$ for Compton
scattering of a photon of 4-momentum $k^{\mu}$
by an electron (of charge $e$ and mass $m_{e}$) with
4-momentum $p^{\mu}$, to a photon momentum $k^{\prime\mu}$ and an
electron momentum $p^{\prime\mu}$ (whose energy is $E'$) is~\cite{ber-quan}:
\begin{eqnarray}
W &=& \frac{(e^{2}/4\pi)^{2} \bar{X}}{2\omega\omega' E E'}
\delta^{4}(p+k-p'-k') \\
\bar{X} &\equiv& 4m_{e}^{4} \left(\frac{1}{\kappa} + \frac{1}{\kappa'}
\right)^{2} - 4m_{e}^{2} \left(\frac{1}{\kappa} + \frac{1}{\kappa'}\right) -
\left(\frac{\kappa}{\kappa'} + \frac{\kappa'}{\kappa}\right),
\end{eqnarray}
with $\kappa \equiv -2p^{\mu}k_{\mu}$ and $\kappa'\equiv
2p^{\mu}k^{\prime}_{\mu}$. 

The electron distribution function is assumed to be a relativistic
Fermi distribution in the frame moving at $\bbeta$. Since we are ignoring
degeneracy effects, in the frame of the radiation we have
\begin{equation}
f(E, \bphat) \approx \et{-[\gamma (E-\bbeta\dt\bp) - \mu_{e}]/T_{e}},
\end{equation}
where $\mu_{e}$ is the electron chemical potential and
$\gamma \equiv (1-\beta^{2})^{-1/2}$. Substituting for
$f(E, \bphat)$ into equation~\eqref{eq_bol}, setting $n(\omega, \bkhat)
=n(\omega)$ in the integrand, and expanding the distribution
functions in powers of $\Delta x$, where
\begin{eqnarray}
x & \equiv & \omega / T_{e}, \\
\Delta x &\equiv & (\omega' - \omega)/T_{e},
\end{eqnarray}
gives the Fokker-Planck expansion for an isotropic radiation
field~\cite{nozawa98}
\begin{eqnarray}
\frac{D n(\omega, \bkhat)}{D t} & = &
2 \left[ \frac{\partial n}{\partial x} I_{1, 0} + n(1+n) I_{1,1} \right]
+  2\left[\frac{\partial^{2} n}{\partial x^{2}} I_{2,0}
+ 2(1+n) \frac{\partial n}{\partial x} I_{2,1} + n(1+n) I_{2,2} \right]
\nonumber \\
&&\mbox{} + 2 \left[ \frac{\partial^{3} n}{\partial x^{3}} I_{3,0}
+ 3(1+n)\frac{\partial^{2} n}{\partial x^{2}} I_{3,1}
+ 3(1+n)\frac{\partial n}{\partial x} I_{3,2}
+ n(1+n)I_{3,3}\right] \nonumber \\
&& \mbox{} + \cdots + 2 n \left[(1+n) J_{0} + \frac{\partial n}{\partial x}
J_{1} + \frac{\partial^{2} n}{\partial x^{2}} J_{2} + \cdots \right],
\label{eq_fok}
\end{eqnarray}
where
\begin{eqnarray}
I_{k,l} & \equiv & \frac{1}{k!} \int \frac{d^{3} \bp}{(2\pi)^{3}}
d^{3} \bp' d^{3} \bk' W f(E, \bphat) (\Delta x)^{k}
\et{x \gamma \bbeta \dt (\bkhat - \bkhat ')} \gamma^{l}
\left(1-\bbeta\dt\bkhat'\right)^{l}, \label{eq_ikl}\\
J_{k} & \equiv & - \frac{1}{k!} \int \frac{d^{3} \bp}{(2\pi)^{3}}
d^{3} \bp' d^{3} \bk' W f(E, \bphat) (\Delta x)^{k}
\left(1- \et{x\gamma \bbeta\dt (\bkhat - \bkhat')}\right). \label{eq_jk}
\end{eqnarray}

We calculate the $I_{k,l}$ and $J_{k}$ coefficients by expanding the
integrands of equations~\eqref{eq_ikl} and~\eqref{eq_jk} in powers of
$p/m_{e}$ and $\omega/m_{e}$. These integrations are ideally suited to
symbolic computer algebra packages (we used Maple). To derive a kinetic
equation correct to $O(\thte^{2}, \beta \thte^{2}, \beta^{2} \thte)$,
one must evaluate $I_{1,0}$ through to $I_{5,5}$, and
$J_{0}$ through to $J_{4}$. Substituting the resulting coefficients back
into equation~\eqref{eq_fok}, we find the kinetic equation
\begin{eqnarray}
\lefteqn{%
\frac{1}{N_{e}\sigma_{T}} \frac{D n(\omega, \bkhat)}{D t} =
\frac{1}{x^{2}} \frac{\partial}{\partial x}\Biggl\{
\thte x^{4} \left[\np + n(1+n)\right] + \thte^{2} \Biggl[
{\frac{5}{2}} x^{4} \left(\np + n(1+n)\right)} \nonumber \\
& & + {\frac{7}{10}} \frac{\partial}
{\partial x} \left( x^{6} \npp{2} \right) + {\frac{7}{5}}x^{3}(1+2n)
\frac{\partial}{\partial x} \left(x^{3} \np\right) + {\frac{7}{10}}
x^{6} \np \left(1-2\np\right) \Biggr] 
+ {\frac{1}{3}} \beta^{2} x^{4} \np \nonumber \\
&&+ \beta^{2} \thte \Biggl[
{\frac{5}{2}} x^{4} \np + {\frac{7}{15}} \frac{\partial}{\partial x}
\left(x^{6} \npp{2}\right) + {\frac{4}{3}} x^{4} n(1+n) + {\frac{7}{15}}
x^{3} (1+2n)\frac{\partial}{\partial x} \left(x^{3} \np\right)\nonumber \\
&& - {\frac{7}{15}}
x^{6} \left(\np\right)^{2} \Biggr] \Biggr\}
- x P_{1}(\mu) \beta \Biggl[\np + \thte C_{1} + \thte^{2} C_{2} \Biggr]
+x P_{2}(\mu) \beta^{2} \Biggl[\frac{2}{3} \np + {\frac{11}{30}} x\npp{2}
+\thte C_{3} \Biggr] \nonumber \\
&&+ O(\thte^{3},\beta
\thte^{3}, \beta^{2} \thte^{2},\beta^{3}),
\label{eq_komp}
\end{eqnarray}
where $\mu$ is the cosine of the angle between the photon momentum and the
peculiar velocity of the electron distribution,
$\mu= \bkhat \dt \bbeta /\beta$,
the $P_{l}(\mu)$ are the Legendre polynomials, and $N_{e}$ is the number
density of electrons in the frame where the bulk velocity vanishes.
The coefficient $C_{1}$ is given by
\begin{equation}
C_{1} = 10\np + \frac{1}{5} x\left(47\npp{2}+7x\npp{3}\right) + 8n(1+n)
+\frac{1}{5}x(1+2n)\left(31 \np + 7 x\npp{2} \right),
\end{equation}
$C_{2}$ by
\begin{eqnarray}
C_{2} &=& 25 \np + {\frac{1}{10}}x \left(1117 \npp{2} + 847 x\npp{3}
+ 183 x^{2}\npp{4} + 11 x^{3} \npp{5} \right)\nonumber \\
&&\mbox{} + 20 n(1+n) + {\frac{1}{10}} x (1+2n)
\left(911\np + 1015 x \npp{2} + 292 x^{2} \npp{3} + 22 x^{3} \npp{4} \right)
\nonumber \\
&& \mbox{}
+ {\frac{1}{10}} x^{2} \left(273 \np + 109 x \npp{2} + 11 x^{2} \npp{3}\right),
\end{eqnarray}
and $C_{3}$ by
\begin{eqnarray}
C_{3} & = & 4 \np + 12 x \npp{2} + 6 x^{2} \npp{3} + {\frac{19}{30}}x^{3}
\npp{4} \nonumber \\
&& \mbox{} + {\frac{8}{3}} n(1+n) + {\frac{1}{30}} x (1+2n) \left(
188 \np + 132 x \npp{2} + 19 x^{2} \npp{3} \right).
\label{eq_c3}
\end{eqnarray}

Equations~\eqref{eq_komp}--\eqref{eq_c3} are the main result of this paper.
Some of the terms in equation~\eqref{eq_komp} have been given previously in
the literature; setting
$\beta=0$ recovers the kinetic equation given in Challinor \&\ Lasenby (1998)
which we used to investigate corrections to the thermal Sunyaev-Zel'dovich
effect (higher-order corrections to the $\beta=0$ equation were given by
Itoh et al. (1998)), the $O(\beta^{2})$ term inside the curly braces
is implicit in the $l=0$ moment equation given in Psaltis \&\ Lamb (1997)
(but not the $O(\beta^{2} \thte)$ term which is fourth-order in the
electron velocity), and the $O(\beta^{2})$ term and part of the
$O(\beta \thte)$ term are implicit in the analysis of Sazonov \&\ Sunyaev
(1998), although they have ignored the parts of these terms arising from
induced scattering and recoil effects. For any particular application,
the validity of neglecting the higher-order terms in equation~\eqref{eq_komp}
should be carefully checked. The terms that we have given are more than
sufficient to describe the kinematic Sunyaev-Zel'dovich effect for typical
cluster parameters, $\beta \simeq 1/300$ and $k_{B}T_{e}\simeq 10\kev$.
(In Nozawa et al.\ (1998) it is shown that the $O(\beta^{2})$ terms are
insignificant for these parameters, while the $O(\beta \thte)$
term gives a correction of $-8.2\%$ to the kinematic Sunyaev-Zel'dovich effect
at the position of the zero of the thermal effect, and the
$O(\beta \thte^{2})$ term gives a correction of $+1.3\%$.)
If required, higher-order terms
in the kinetic equation can be derived from the Fokker-Planck expansion
(eq.~\eqref{eq_fok}), although in practice the evaluation of the $I_{k,l}$
and $J_{k}$ rapidly becomes prohibitive.

We have written equation~\eqref{eq_komp} in a form that manifestly
preserves the total number of photons, as required for Compton scattering.
For $\beta=0$, we obtain a generalisation of the diffusion approximation
to the Boltzmann equation (see, for example, Prasad et al.\ (1988)).
Note that we have derived our equation (eq.~\eqref{eq_komp}) by a systematic
expansion of the original Boltzmann equation (eq.~\eqref{eq_bol}) in
$1/m_{e}$; we have not appealed to the heuristic arguments that
form the basis of the diffusion approximation. 
For a spatially localised, isotropic distribution of photons, the
time rate of change of the total photon number $N_{r}$ is given by
\begin{equation}
\frac{d N_{r}}{d t} = 2 \int \frac{d^{3} \bk}{(2\pi)^{3}}
d^{3} \bx \frac{D n(\omega, \bkhat)}{D t},
\end{equation}
where $d^{3}\bx$ is the spatial measure, and the factor of two accounts for
the two polarisations.
Integrating equation~\eqref{eq_komp}
over photon momenta, the terms involving $\mu$ vanish by virtue of the
integral over solid angles, and the $\mu$-independent terms (those
in curly braces) vanish after integration over photon energies, since these
terms are written in the form of a conservation law. It is not hard
to show that equation~\eqref{eq_komp} with $\beta=0$ admits static,
homogeneous solutions with $n = 1/(\exp(x-\nu)-1)$ as required for
thermodynamic equilibrium.

\section{Rate of energy transfer}

The rate of increase of energy density $E_{r}$ in the radiation
due to Compton scattering is given by
\begin{equation}
\frac{\partial E_{r}}{\partial t} = 2 \int \frac{d^{3} \bk}{(2\pi)^{3}}
\frac{D n(\omega,\bkhat)}{D t} \omega.
\end{equation}
Substituting for $Dn/Dt$ from equation~\eqref{eq_komp}, we see that only the
$\mu$-independent terms contribute to the energy transfer. Substituting a
Planck distribution at temperature $T_{r}$ for $n$ and performing the
integral, we find
\begin{eqnarray}
\frac{dE_{r}}{dt} &=& E_{r} N_{e} \sigma_{T}\biggl\{
4(\thte-\thtr)\left[1+{\frac{5}{2}}\thte - 21 {\frac{\zeta(6)}{\zeta(4)}}
\thtr + O(\theta^{2}) \right] \nonumber
\\
&&\mbox{}+ \beta^{2} \left[ {\frac{4}{3}} + 10 \thte
-\left({\frac{16}{3}}+28 {\frac{\zeta(6)}{\zeta(4)}} \right)\thtr
+ O(\theta^2) \right] + O(\beta^{4}) \biggr\},
\label{eq_erg}
\end{eqnarray}
where $\zeta(x)$ is the Riemann Zeta function, and $\thtr \equiv T_{r}/m_{e}$.
The terms in the first square bracket in equation~\eqref{eq_erg} are
independent of $\beta$; they tend to equalise the radiation and electron
temperatures. These terms (which were also given in Challinor \&\ Lasenby
(1998)) were first derived by Woodward (1970), where higher-order terms were
also given. The terms in the second square bracket in equation~\eqref{eq_erg}
represent the lowest-order effects of the electron bulk velocity on the
energy transfer. The first such term $4E_{r}\sigma_{T} N_{e}\beta^{2} /3$
is well known (see, for example, Sazonov \&\ Sunyaev (1998) and references
within).

\section{The Sunyaev-Zel'dovich effect}

For CMB photons passing through a cluster at redshift $z$, the average
value of $x=\omega/T_{e}$ is $\bar{x} \simeq 6.2 \times 10^{-4} (1+z)/
k_{B} T_{e}$, where $k_{B} T_{e}$ is expressed in $\ev$. Since the electron
temperature for a hot cluster is typically $\simeq 10\kev$, it follows that
$\bar{x} \ll 1$. In this limit, equation~\eqref{eq_komp} reduces to
\begin{eqnarray}
\lefteqn{%
\frac{1}{N_{e}\sigma_{T}} \frac{D n(\omega, \bkhat)}{D t} =
\frac{1}{x^{2}} \frac{\partial}{\partial x}\Biggl\{
x^{4}\left[\thte + \frac{1}{3}\beta^{2}+\frac{5}{2}\thte(\thte+\beta^{2})
\right]\np} \nonumber \\ &&\mbox{}+
\thte\left[\frac{7}{10}\thte+\frac{7}{15}\beta^{2}\right]
\frac{\partial}{\partial x}\left(x^{6}\npp{2}\right)\Biggr\}
- x P_{1}(\mu)\beta \Biggl[ \np + \thte \left(10\np + \frac{47}{5}
x\npp{2} + \frac{7}{5} x^{2}\npp{3} \right)\nonumber \\
&&\mbox{}+ \thte^{2} \left(25\np +
\frac{1117}{10}x\npp{2} + \frac{847}{10}x^{2}\npp{3} + \frac{183}{10}
x^{3}\npp{4} + \frac{11}{10} x^{4} \npp{5} \right) \Biggr] \nonumber \\
&&\mbox{}+x P_{2}(\mu)\beta^{2} \Biggl[\frac{2}{3} \np + \frac{11}{30} x\npp{2}
+\thte\left(4\np+12x\npp{2} + 6x^{2}\npp{3} + \frac{19}{30}x^{3} \npp{4}
\right) \Biggr].
\end{eqnarray}
Setting $n=1/(\exp(\alpha x)-1)$, where $\alpha \equiv T_{e}/T_{r}$
recovers the combined thermal and kinematic Sunyaev-Zel'dovich spectral
distortion given in Nozawa et al.\ (1998) and Sazonov \&\ Sunyaev (1998),
where the relative importance of the various terms for typical cluster
parameters is discussed in detail.

\section{Conclusion}

Using the covariant Fokker-Planck expansion described in Nozawa et al.\ (1998),
we have derived a kinetic equation describing the interaction of an
isotropic radiation field with a thermal distribution of electrons, which moves
at bulk velocity $c\beta$ relative to the radiation, in the limit of
low optical depth. Relativistic effects are
included to $O(\thte^{2}, \beta\thte^{2}, \beta^{2}\thte)$, or equivalently
to $O(\Theta^{2}, \beta\Theta^{2}, \beta^{2}\Theta)$ where $\Theta$ is
either of $\hbar \omega/ m_{e} c^{2}$ or $k_{B} T_{e} / m_{e} c^{2}$, which
is sufficient to describe the corrections to the kinematic Sunyaev-Zel'dovich
effect for typical cluster parameters (Nozawa et al. 1998)
The method may be easily extended to include higher-order relativistic
effects if required. We have calculated the rate of energy transfer between
a Planckian radiation field and the electrons, obtaining the usual
``thermal'' and ``kinematic'' terms, as well $O(\beta^{2} \theta)$
``interference'' terms. Specialising to the limit $T_{r} \ll T_{e}$, we
confirm the relativistic corrections to the thermal and kinematic
Sunyaev-Zel'dovich effect given in Nozawa et al.\ (1998) and
Sazonov \&\ Sunyaev (1998).

\acknowledgements

We would like to express our gratitude to Roberto Turolla and Silvia Zane
for bringing a number of useful references to our attention.


\begin{thebibliography}{10}

\bibitem[(Berestetskii, Lifshitz, \&\ Pitaevskii 1982)]{ber-quan}
Berestetskii V. B., Lifshitz E. M., \&\ Pitaevskii L. P. 1982,
Quantum Electrodynamics: Landau and Lifshitz Course of Theoretical Physics,
second edition (Oxford: Pergamon Press plc)

\bibitem[(Buchler \&\ Yeuh 1976)]{buchler76}
Buchler J. R., \&\ Yeuh W. R. 1976, \apj, 210, 440

\bibitem[(Challinor \&\ Lasenby 1998)]{chall98}
Challinor, A. D., \&\ Lasenby, A. N. 1998, Relativistic corrections to the
Sunyaev-Zel'dovich effect, \apj, in press

\bibitem[(Itoh et al.\ 1998)]{itoh98}
Itoh, N., Kohyama, Y., \&\ Nozawa, S. 1998, Relativistic corrections to the
Sunyaev-Zel'dovich effect for clusters of galaxies, \apj, in press

\bibitem[(Kershaw, Prasad, \&\ Beason 1986)]{kershaw86}
Kershaw, D. S., Prasad, M. K., \&\ Beason, J. D. 1986,
J. Quantit. Spectrosc. Radiat. Transfer, 36, 273

\bibitem[(Kompaneets 1957)]{komp57}
Kompaneets, A. S. 1957, Soviet Physics JETP, 4, 730

\bibitem[(Nozawa et al.\ 1998)]{nozawa98}
Nozawa, S., Itoh, N., \&\ Kohyama, Y. 1998, Relativistic corrections to the
Sunyaev-Zel'dovich effect for clusters of galaxies. II. Inclusion of peculiar
velocities, submitted

\bibitem[(Peebles \&\ Yu 1970)]{peebles70}
Peebles, P. J. E., \&\ Yu, J. T. 1970, \apj, 162, 815

\bibitem[(Prasad et al.\ 1988)]{prasad88}
Prasad, M. K., Shestakov, A. I., Kershaw, D. S., \&\ Zimmerman, G. B.
1988, J. Quantit. Spectrosc. Radiat. Transfer, 40, 29

\bibitem[(Psaltis \&\ Lamb 1997)]{psaltis97}
Psaltis, D. \&\ Lamb, F. K. 1997, \apj, 488, 881

\bibitem[(Rephaeli 1995)]{reph95}
Rephaeli, Y. 1995, \apj, 445, 33

\bibitem[(Rephaeli \&\ Yankovitch 1997)]{reph97}
Rephaeli, Y., \&\ Yankovitch, D. 1997, \apj, 481, L55

\bibitem[(Sazonov \&\ Sunyaev 1998)]{sazonov98}
Sazonov, S. Y., \&\ Sunyaev, R. A. 1998, Cosmic microwave background radiation
in the direction of a moving cluster of galaxies with hot gas: relativistic
corrections, submitted

\bibitem[(Stebbins 1998)]{stebbins98}
Stebbins, A. 1998, Extensions to the Kompaneets equation and
Sunyaev-Zel'dovich distortion, \apj, in press

\bibitem[(Woodward 1970)]{woodward70}
Woodward, P. 1970, \prd, 1, 2731

\bibitem[(Zel'dovich \&\ Sunyaev 1969)]{zel69}
Zel'dovich, Ya. B., \&\ Sunyaev, R. A. 1969, \apss, 4, 301

\end{thebibliography}
\end{document}